\documentclass[final,5p,times,twocolumn]{elsarticle}

\usepackage{graphicx,subfigure,float}% Include figure files
\graphicspath{{Figures/}}
\usepackage{lineno,hyperref,xcolor,url}
\hypersetup{colorlinks=true,citecolor=blue%,urlcolor=cyan
}
\biboptions{sort&compress}
\begin{document}

\begin{frontmatter}

\title{Stencil growth of metallic nanorod: An atomistic simulation%\tnoteref{mytitlenote}
}
\author{Movaffaq Kateb}\corref{cor1}
\ead{movaffaqk@ru.is}
%\author[2,3]{Jon Tomas Gudmundsson}
%\author[3]{Snorri Ingvarsson}
%\author[1]{Andrei Manolescu}
\address{Department of Engineering, School of Technology, Reykjavik University, Menntavegur 1, IS-102 Reykjavik, Iceland}
%\address[2]{Science Institute, University of Iceland,
%Dunhaga 3, IS-107 Reykjavik, Iceland}
%\address[3]{Department of Space and Plasma Physics, School of Electrical Engineering and Computer Science, \\KTH Royal Institute of Technology, SE-100 44, Stockholm, Sweden}
\cortext[cor1]{Corresponding author}

\begin{abstract}
The stencil growth of nanoscale patterns using molecular dynamic simulation has been demonstrated. A comparison has been made to a film grown by identical conditions without the stencil. It is shown that in the case of nanoscale proximity between mask and substrate, patterns of the same dimension as the mask can be obtained. The results also indicate that the obtained nanorod presents a higher surface area than the corresponding thin film. It is demonstrated that nanorod surface roughness decreases by merging adjacent surface irregularity during the deposition.
\end{abstract}


\end{frontmatter}

%\linenumbers

\section{Introduction}
Stencil growth is a large area method of pattering wherein a mask is utilized to block a flux of atoms or ions \citep{vazquez2015}. In particular, it is advantageous when adhesion of metallic nanostructures to the substrate is poor and alternative patterning methods may detach nanostructures. It allows growing various nanostructures such as nanodots \citep{champagne2003} and nanowires \citep{wasserman2008} without resist i.e.\ avoiding contamination from organic materials and solvents, radiation or applying pressure. The pattern obtained by stencil deposition retains the nominal roughness of as-deposited metal films which gives enhanced propagation length for surface plasmon polaritons \citep{lindquist2012}. 

A huge effort has been made in order to increase the durability of the stencil, reduction of the resulting feature size and reproducibility of deposited pattern \citep{vazquez2015}. For instance, it has been shown that deposition flux through apertures as small as 5\,nm is feasible \citep{deshmukh1999}. However, the resulting pattern was larger due to the distance between the mask and substrate. A small feature size also requires a collimated deposition flux which is not achieved in e.g.\ sputtering \citep{lindquist2012}. While the method has already been established for wafer scale sub-micron patterns (down to 100\,nm), further improvements in terms of resolution and surface roughness require enormous experimental effort and complicated characterization techniques.

In this regard, molecular dynamics (MD) simulation has shown promising owing to its atomistic resolution. It has been utilized for revealing atomistic mechanisms that contribute to various physical vapor deposition (PVD) methods \citep{kateb2019,kateb2020}. For instance, using MD simulation it is observed that Cu films obtained by different PVD methods suffering from the presence of twinning \citep{kateb2019} which has been verified by experimental studies \citep{chen2013,cemin17:120}. Besides, the dynamic nature of MD simulation allowed to understand the mechanism leading to lower surface roughness in ionized PVD methods.

Despite this great potential, MD simulation of stencil growth has not been studied so far. Besides, most of the previous studies i.e.\ depositions without stencil were suffering from unrealistic assumptions e.g.\ being performed in two dimensions, assuming mono-dispersed energy flux, deposition of a small number of atoms and finally neglecting ions (the interested reader is referred to Ref.\ \citep{kateb2019} and references therein).  

In the present study, we demonstrate the thermal evaporation of Cu with and without the stencil using MD simulation. The evaporation without stencil helps to distinguish between intrinsic characteristics of evaporated film and those dictated by the stencil. Special attention has been brought to the pattern size and lower surface roughness obtained during the stencil deposition.

%----------------------------------------------------------------------------
\section{Method}

MD simulations \citep{allen1989} were performed using the LAMMPS \citep{plimpton1995,plimpton2012} package\footnote{LAMMPS website, \url{http://lammps.sandia.gov/}, distribution 14-April-2018}. We assumed a fully neutral flux to represent thermal evaporation \citep{kateb2019}. The embedded-atom method (EAM) force field \citep{daw1983,daw1984} was employed to model the interactions of film/substrate atoms. The substrate was a single crystal Cu consist of 16 (111) planes with 77$\times$90\,{\AA}$^2$ dimensions. This makes the $\langle111\rangle$ orientation to be the growth direction. The substrate was divided into 3-layer as proposed by \citet{srivastava1989}. A fixed monolayer at the bottom to prevent the rest from moving after collisions, then three monolayers as a thermostat layer to control heat dissipation and prevents melting and the rest of the substrate as a surface layer. The initial velocities of substrate atoms were defined randomly from a Gaussian distribution to mimic the temperature of 300\,K and the substrate energy was minimized prior to relaxation.

For both cases, the deposition flux consisting of 22000 atoms was introduced at a distance of 15\,nm above the substrate surface. For the stencil growth, the incoming flux was confined in a circle of 6\,nm in diameter in the middle of the substrate. Thus, we are modeling the final stage growth where flux of atoms passed the mask. For deposition without stencil flux of atoms was introduced everywhere above the substrate. The initial velocity of ions towards the substrate were assigned randomly with a flat distribution within 0\,--\,10\,eV energy range. 
The process of introducing species was a single atom/ion every 0.1\,ps which produces an equal deposition rate in both cases. One may think this will generate a deposition rate several orders of magnitude larger than that of a typical experiment. We have recently shown that even a higher deposition rate \citep{kateb2019} gives a similar result to that of accelerated MD simulation \citep{hubartt2013} with deposition rate similar to that of experimental studies.

The Verlet \citep{kateb2012} algorithm was used for time integration of the equation of motion with a timestep of 5\,fs by sampleing from microcanonical (NVE) ensemble. The thermostat layer was controlled by Langevin thermostat \citep{schneider1978} with a damping of 50\,fs for a total time of 25\,ns. The damping defines the timescale for resetting temperature and that generates a heat flow towards the thermostat layer.

We utilized common neighbor analysis (CNA)  \citep{steinhardt1983,kelchner1998,tsuzuki2007} to characterize the local structure. CNA allows distinction between fcc and hcp which is of practical importance in defect analysis  \citep{kateb2018b,azadeh2019}. The Ovito package\footnote{Ovito website, \url{http://ovito.org/}, Version 3.0.0-dev362} was used to generate atomistic illustrations and post-processing of CNA. \citep{stukowski2009}

\section{Results \& discussion}
\subsection{Surface roughness}
\begin{figure}[h]
    \centering
    \includegraphics[width=1\linewidth]{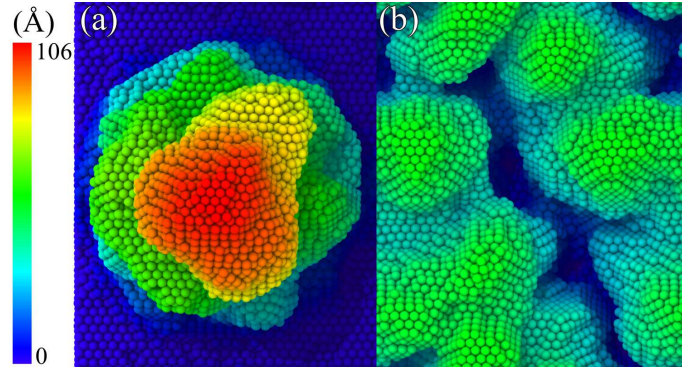}
    \caption{The surface roughness of films deposited by evaporation (a) with and (b) without the stencil. The colorbar indicates height from the original substrate surface.}
    \label{fig:z}
\end{figure}

Fig.~\ref{fig:z} shows the final topology of nanorod and film. The colorbar is used to illustrate the height where dark blue indicates the original substrate surface ($Z=0$) and red denotes the height of 10.6\,nm above it. It can be seen that using stencil the height topmost film atom ($Z_{\rm film}^{\rm max}$) is twice that without masking (6\,nm). Note that the same number of adatoms deposited in both cases without any resputtering from the surface (substrate and later film).

The summary of different values are presented in Table~\ref{tab:z}. The lowermost film atoms indicated by $Z_{\rm film}^{\rm min}$ have negative $Z$ which means they are embedded into the substrate. The latter values must be compared to the topmost substrate atom denoted by $Z_{\rm sub}^{\rm max}$ to give an estimation of intermixing. The difference is negligible but it seems stencil causes more interface mixing and thus stronger adhesion to the substrate. Note that identical energy distribution is utilized in both cases for defining initial velocity of adatoms. Thus, the larger interface mixing with the stencil can be attributed to a more localized bombardment by adatoms. The surface area ($A$) of the nanorod obtained by stencil is higher than the film. We also compared the volume change before and after deposition ($\Delta V$). It can be seen that slightly higher volume obtained in the uniform deposition that is an indication of closed pores.

\begin{table}[h]
\centering
\caption{\label{tab:z} $A$ denote film surface area, $\Delta V$ refers to volume change before and after sputtering and $Z$ indicates maximum/minimum position of the film/substrate atoms with respect to the initial substrate surface.}
\begin{tabular}{ c c c }
\hline \hline
Method & with & without \\
\hline
%$N$ (atom) & 22000 & 21997 \\
$Z_{\rm film}^{\rm min}$ ({\AA}) & -10.60 & -4.34\\
$Z_{\rm film}^{\rm max}$ ({\AA}) & 105.72 & 60.05\\
$Z_{\rm sub}^{\rm max}$ ({\AA}) & 6.09 & 1.95 \\
$A$ (nm$^2$) & 232.704 & 176.839 \\
$\Delta V$ (nm$^3$) & 248.924 & 255.735 \\
\hline
\end{tabular}
\end{table}

\subsection{Microstructure}
\begin{figure}[h]
    \centering
    \includegraphics[width=1\linewidth]{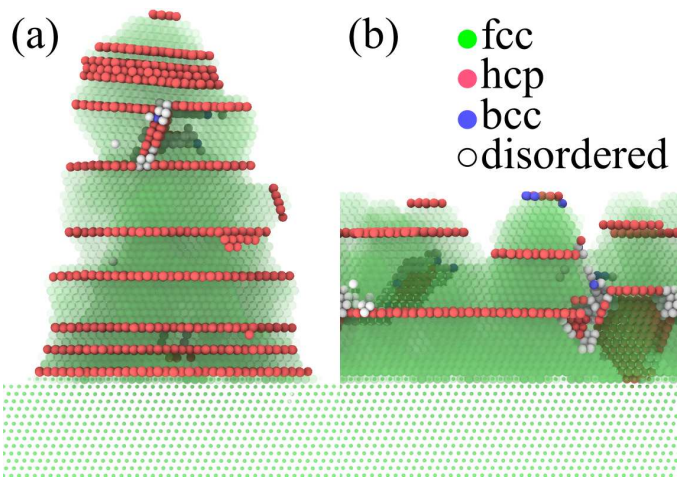}
    \caption{Demonstration of microstructure using CNA for deposition (a) with and (b) without the stencil. The surface atoms were removed and fcc atoms are shown semi-transparent for illustration of internal defects. The substrate atoms are illustrated with a smaller diameter.}
    \label{fig:cna}
\end{figure}

The final microstructure of films deposited with and without stencil is shown in Fig.~\ref{fig:cna}. The surface atoms were removed and fcc atoms are shown semitransparent for illustration of internal defects. It is evident that there are more hcp atoms in the nanorod while more internal defects exist in the thin film. The confinement of deposition flux by the stencil simulates a higher deposition rate. As a result a more non-equilibrium growth, hcp phase, is expected with the stencil. We would like to remark that the hcp phase existence in an fcc crystal can be translated into stacking fault areas and twining. This has been observed experimentally using the X-ray diffraction pole figures of an evaporated Cu film \citep{chen2013}. Besides accelerated MD simulation with realistic deposition rate verified the existence of hcp phases in evaporated Cu \citep{hubartt2013}.

\begin{figure}[h]
    \centering
    \includegraphics[width=1\linewidth]{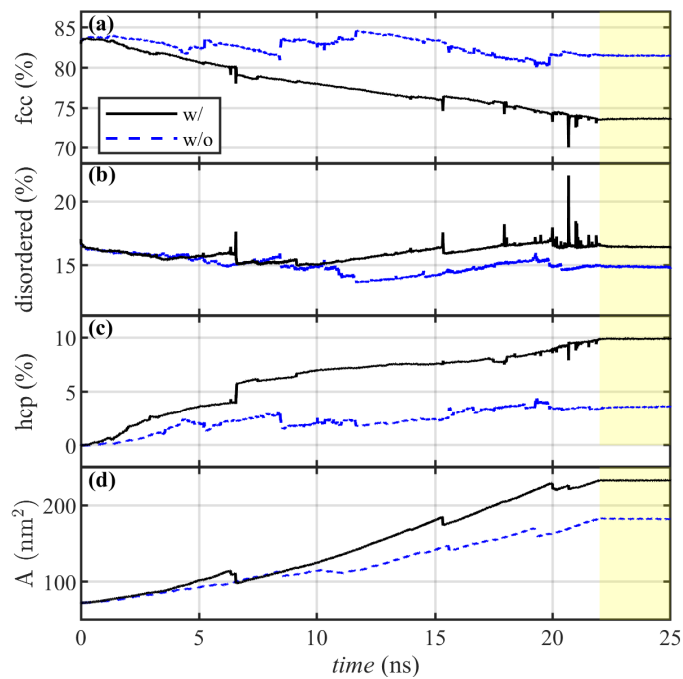}
    \caption{Time evolution of (a) fcc, (b) disordered, (c) hcp and (d) surface area for deposition with and without the stencil. The highlighted part (22\,--\,25\,ns) indicate post-deposition relaxation.}
    \label{fig:cna-time}
\end{figure}

Fig.~\ref{fig:cna-time} shows the time evolution of the microstructure and $A$ during deposition. In both cases, there is a decreasing trend in the fcc ratio while the hcp ratio shows an increasing trend. However, these changes are more pronounced in the case of deposition with the stencil. Since fcc is the equilibrium structure of copper, more increase in hcp and drop in fcc ratio indicate increased defects density. The disordered ratio remains almost constant but presents smaller value for deposition without stencil. It is worth mentioning that surface atoms are characterized as disordered by CNA due to the lack of symmetry. Thus lower disordered ratio means lower surface to volume ratios as indicated in table~\ref{tab:z}. 

\begin{figure}
    \centering
    \includegraphics[width=1\linewidth]{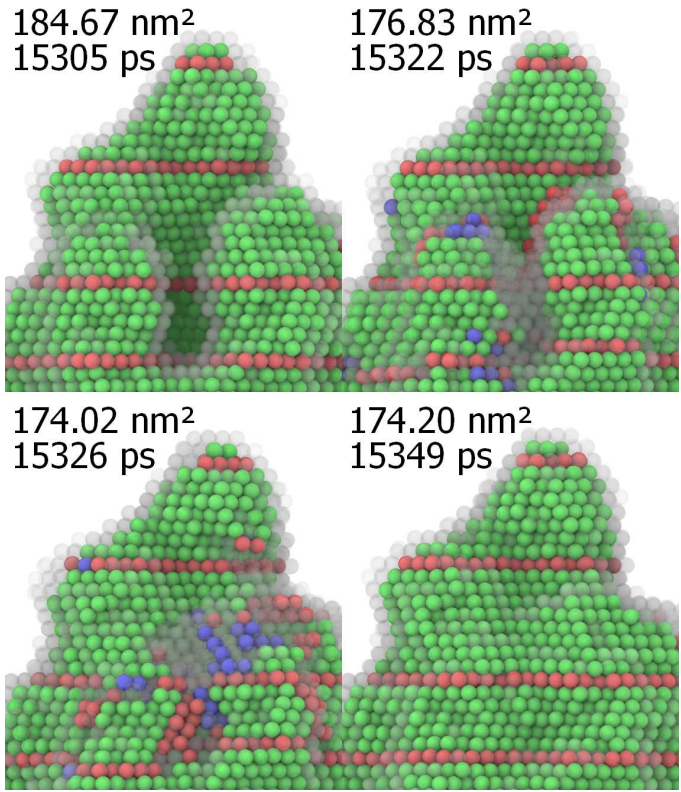}
    \caption{Snapshots of the merging that indicate a reduction of 10\,nm$^2$ in the surface area through the formation of a disordered phase (smoky atoms) between surface irregularities. Legends are indicating surface area and time.}
    \label{fig:merge}
\end{figure}

It can be seen that with stencil there are sharp peaks in disordered structure ratio (Fig.~\ref{fig:cna-time}b). These peaks indicate transitions e.g.\ merging of two adjacent islands into one. An example of the merging is illustrated in Fig.~\ref{fig:merge} which indicate a relatively sharp valley becomes unstable and vanishes. Using CNA it can be detected by an increase in the ratio of disordered atoms (15322\,ps) that immediately followed by nucleation of fcc or hcp phase (15326\,ps). As a result, the ratio of disordered atoms is reduced which is associated with a drop in $A$ (Fig.~\ref{fig:cna-time}d). This mechanism clearly explains the reduced surface roughness obtained by stencil deposition. The highlighted area in the figure (after 22\,ns) indicates that there is no change in these quantities after deposition. Thus, it can be concluded that the merging mechanism is associated with bombarding adatoms.

\section{Summary}
In summary, we demonstrated the stencil deposition using MD simulation. It has been shown that close proximity of the mask-substrate allows growing patterns (nanorod here) of the same lateral dimensions of the mask. Besides, the results indicate lower surface roughness for nanorod obtained by the stencil deposition compared to the corresponding thin film. This is achieved through the merging mechanism of adjacent surface irregularities induced by low energy bombardment.

%\section*{acknowledgements}
%This work was partially supported by the University of Iceland Research Funds and the Icelandic Research Fund Grant Nos.~195943-051, 196141, 130029 and 120002023.

\bibliography{ref}

\end{document}